\documentclass[3p,twocolumn]{elsarticle}
\usepackage{amssymb}
\journal{Solid State Communication}

\begin{document}

\begin{frontmatter}

\title{{Eccentricity effects on the quantum confinement in double quantum rings}}

\author[label1]{J. Costa e Silva}
\author[label2]{A. Chaves \corref{label5}}\ead{andrey@fisica.ufc.br}
\author[label2]{G. A. Farias}
\author[label3]{M. H. Degani}
\author[label4]{R. Ferreira}
\address[label1]{Universidade Federal Rural do Semi-\'Arido, Campus
Central Costa e Silva, CEP 59600-900, Mossor\'o, RN, Brazil}
\address[label2]{Departamento de F\'{\i}sica, Universidade Federal do
Cear\'a, Caixa Postal 6030, 60455-900, Fortaleza, CE, Brazil}
\address[label3]{Haras Degani, Av. Fioravante Piovani, 1000 Jardim das
Laranjeiras, CEP 13257-700, Itatiba, SP, Brazil}
\address[label4]{Laboratoire Pierre Aigrain, Ecole Normale Superieure,
CNRS UMR 8551, Universit\'e P. et M. Curie, Universit\'e Paris
Diderot, 24 rue Lhomond, F-75005 Paris, France}
\cortext[label5]{Corresponding author: Tel.: +55 85 33669018; Fax: +55 85 33669450}

\begin{abstract}
This work presents a theoretical study of the energy spectrum of
GaAs/AlGaAs concentric double quantum rings, under an applied
magnetic field directed perpendicular to the ring plane. The
Schr\"odinger equation for this system is solved in a realistic
model consisting of rings with finite barrier potentials.
Numerical results show that increasing the magnetic field
intensity leads to oscillations in the ground state energy which,
in contrast to the usual Aharonov-Bohm oscillations, do not have a
well defined period, due to the coupling between inner and
outer rings states. However, when one considers an elliptical
geometry for the rings, the energy spectra of the inner and
outer ring states are decoupled and the periodicity of the
oscillations is recovered.
\end{abstract}

\end{frontmatter}

\section{Introduction}

The increasing development of growth techniques has made it
possible to fabricate nanometric structures which exhibit strong
quantum confinement, such as quantum dots\cite{Ledentsov, Skolnick} and quantum rings
\cite{Kleemans, Warburton, 7}. Semiconductor quantum rings (QRs) have attracted much
attention because they combine the optical properties of
self-assembled nanostructures with unique features under applied
magnetic fields originated by the ring topology, namely the
Aharonov-Bohm (AB) effect \cite{Aharonov, Wendler, 1,6}. Recently, the growth of
self-assembled GaAs/AlGaAs concentric double quantum rings has
been reported, opening a new route to study AB oscillations and
quantum interference effects \cite{5,2}. The electronic structure and excitonic properties of such systems were characterized by photoluminescence measurements in the absence of magnetic fields.\cite{2, Sanguinetti} Physical understanding
and accurate modelling of such double rings are required for their future
design and application, which stimulated the development of recent
theoretical studies on circular concentric double quantum rings.
\cite{Latge1, Latge2, 11, Latge3, Planelles2, Vanska}

Previous theoretical works on semiconductor QRs have demonstrated
interesting phenomena coming from non-circular geometries. The
energy spectrum of a single QR with arbitrary shape was studied by
Gridin et al. \cite{Gridin}, where they observed the existence of
eigenstates localized at regions of maximum curvature. Later,
Bruno-Alfonso and Latg\'e \cite{Alfonso} studied the more general
case of non-uniform ring widths and arbitrary shapes of the
centerline, where it was shown that even tiny variations of the
ring width are able to produce quenching of the AB oscillations in
the lowest energy levels. The specific case of elliptic single QRs
was also considered by previous works, \cite{Gil, Planelles} where
it was shown that the low lying energy states are confined at the
two regions of maximum curvature and their energies group in pairs
exhibiting AB oscillations.

In this work, a time evolution method is used to calculate the
energy eigenstates of electrons in GaAs/AlGaAs concentric double
quantum rings, within the effective mass approximation. We will
consider a model with finite barrier potentials, which is not
limited to small perturbations and which allows us to study rings
with arbitrary sizes and shapes. For circular rings, our numerical
results show that when the magnetic field is increased, energy
oscillations are found, but they do not have a well defined
periodicity, contrary to the AB oscillations found for single
rings in the literature \cite{3,8}. Our results also show that the
periodicity of the AB oscillations is recovered when an elliptic
geometry is considered for one of the rings. In this case, the
periodicity of the oscillations of the energy of the ground and
first excited states is different from that found for the second
and third excited states.
\begin{figure}[!h]
\includegraphics[width= 0.7\linewidth]{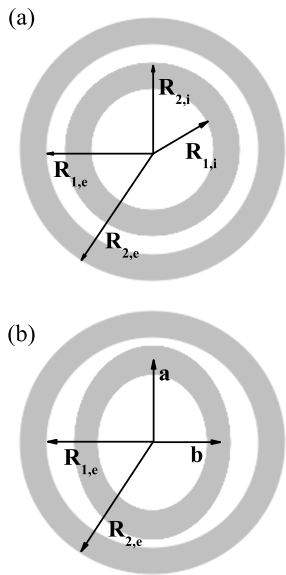}
\caption{Double ring potentials, considering (a) circular and (b)
elliptic inner rings, and circular outer rings.}
\label{fig:1}
\end{figure}

\section{Theoretical Model}

Our model consists of two concentric planar quantum rings, under
an applied magnetic field $\overrightarrow{B} = B \hat{z}$,
\emph{i.e.}, perpendicular to the ring plane. The Hamiltonian for
such a system, using the symmetric gauge $\overrightarrow{A} =
(-By/2,Bx/2,0)$, is given by
\begin{equation}
\label{eq1} H(\overrightarrow{r}) = T_x + T_y + V(x,y) \,,
\end{equation}
where
\begin{equation}
\label{eq2} T_x = \frac{1}{{2m}}\left( { - i\hbar \frac{\partial
}{{\partial x}} - y\frac{eB}{2}} \right)^2 \,,
\end{equation}
and
\begin{equation}
\label{eq3} T_y = \frac{1}{{2m}}\left( { - i\hbar \frac{\partial
}{{\partial y}} + x\frac{eB}{2}} \right)^2 \,,
\end{equation}
are the kinetic energy operators. The confinement potential is
defined as $V(x,y) = 0$ within the ring region and $V(x,y) =
V_{e}$ at the barrier region, where $V_{e}$ is the electron band
offset. Figure 1 illustrates the two cases of double ring
potentials discussed in this work: in Fig. 1(a), both inner and
outer rings are circular, limited by radii $R_{1,i}$ and
$R_{2,i}$ for the inner, and $R_{1,e}$ and $R_{2,e}$ for the
outer rings. In Fig. 1(b), the circular outer ring is kept,
but the inner ring is elliptic. The ellipses are defined by the
eccentricity $\xi = a/b$, where $a$ and $b$ are the average
distance between the limits of the inner ring in $x$ and $y$
directions, respectively, chosen to preserve the inner ring
area.

Notice that in this simple model, the rings are assumed to be planar and spatially separated by a layer of a different material, represented by the potential barrier $V_e$ between them. In fact, the actual self-assembled double quantum rings reported in the literature have a finite height and are not completely spatially separated, namely, they are connected by their bottom. \cite{2, Sanguinetti} However, as the rings height is usually much smaller than their width, we assume that the confined electron is always in the ground state for the vertical confinement, so that the electronic transitions we will study in this work, related to the in-plane confinement, occur for energies lower than the first excited state energy of the vertical direction. As for the bottom connection, we point out that such a connection is a region of lower height separating the inner and outer rings.  This narrowing in the vertical direction enhances the energy in the region between the rings and, consequently, the inner and outer rings are still effectively separated by a higher energy (barrier) region. Hence, we believe that our model, however simple it may be, still captures the most relevant physical aspects of the problem, at least in a qualitative way.

From the time dependent Schr\"odinger equation
\begin{equation}
\label{eq4} i\hbar \frac{\partial}{\partial
t}\Psi(\overrightarrow{r},t) = H \Psi(\overrightarrow{r},t) \,,
\end{equation}
it is straightfoward to find the time evolution of an arbitrary
wavefunction $\Psi(\overrightarrow{r},t)$ as
\begin{equation}
\label{eq5} \Psi(\overrightarrow{r},t+\delta t) = \exp
\left[\frac{i}{\hbar}H\delta t \right]\Psi(\overrightarrow{r},t)
\,.
\end{equation}

The eigenstates of the Hamiltonian of Eq. (\ref{eq1}) are then
found by propagation of an arbitrary initial state in imaginary
time domain $\tau = it$. The time evolution is performed by means
of the split-operator technique. \cite{Gil, meu} The differential
operators in Eq. (\ref{eq2}) and (\ref{eq3}) are discretized,
following a finite differences scheme. When $\tau \rightarrow
\infty$, the initial state converges to the ground state of the
system. Excited states are found by mixing this procedure with the
Gram-Schmidt orthonormalization method. \cite{4}
\begin{figure}[!b]
\includegraphics[width= 0.9\linewidth]{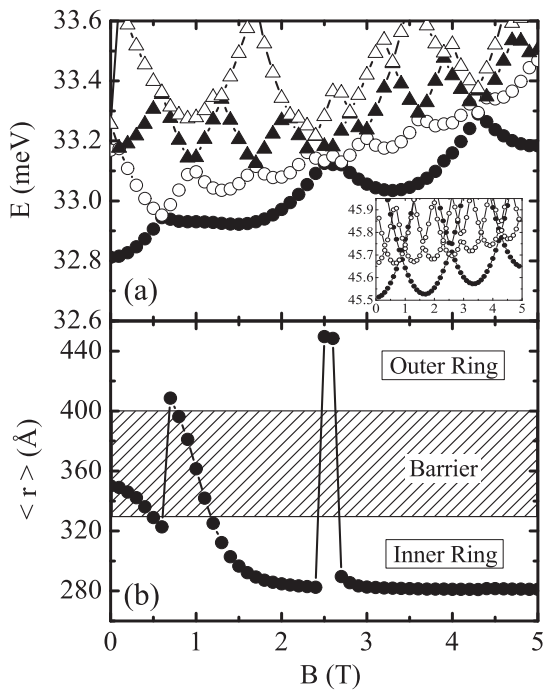}
\caption{(a) Energy spectrum and (b) ground state average radius
$\langle r \rangle = \langle \Psi_0 | r |\Psi_0\rangle$ as a
function of the magnetic field for electrons in double circular
rings with $R_{1,i} = 230$ \AA\,, $R_{2,i} = 330$ \AA\,, $R_{1,e}
= 400$ \AA\, and $R_{2,e} = 500$ \AA\,. Inset: energy spectrum for the case of negligible coupling between the rings, namely, with infinite potential barriers.} \label{fig:2}
\end{figure}

A qualitative analysis of the AB oscillations of the energy
spectra of quantum rings can be made by considering a thin
circular single quantum ring. In this case, neglecting any radial
or perpendicular movement of the electron, i.e. considering that
the electron moves only in the angular $\theta$-direction, with
fixed position $\rho = R$ and $z = 0$ (cylindrical coordinates)
the Hamiltonian can be written as
\begin{equation}
H^{(\theta)}
=\frac{\hbar^2}{2mR^2}\left[-i\frac{d}{d\theta}+\frac{\phi}{\phi_0}\right]^2.
\end{equation}
where $\phi = \pi R^2 B$ is the magnetic flux threading the ring
and $\phi_0 = h/e$ is the quantum flux. The eigenfunctions of this
Hamiltonian are easily found as $\psi_n(\theta) =
\exp(in\theta)/\sqrt{2\pi}$, where $n = 0, \pm 1, \pm 2, ...$ is
the angular momentum index, and the eigenenergies are
\begin{equation}\label{fig:ABspectrum}
E^{(\theta)}_n(\phi)
=\frac{\hbar^2}{2mR^2}\left[n+\frac{\phi}{\phi_0}\right]^2.
\end{equation}
The ground state energy as a function of $\phi$ given by Eq.
(\ref{fig:ABspectrum}) exhibits AB oscillations with period
$\phi_0$ and its angular momentum index changes periodically at
magnetic flux $\phi = (n+1/2)\phi_0$. The results obtained by this
simple approximation show that the period of the AB oscillations
for an electron confined by a ring shaped potential of radius $R$
is $\Delta B = \phi_0\big/\pi R^2$, i.e. it is inversely
proportional to the ring area. \cite{10} This information will be
useful for understanding the main features observed in the energy
spectrum of double quantum rings in the following section.

\section{Results and Discussion}

We have calculated the energy spectrum for electrons in
GaAs/Al$_{0.30}$Ga$_{0.70}$As double concentric quantum rings. For
such a heterostructure, the electron band offset is $V_e = 262$
meV and the effective mass is assumed as $m_e = 0.067 m_0$
\cite{2}.
\begin{figure} [!h]
\centering
\includegraphics[width= 0.9\linewidth]{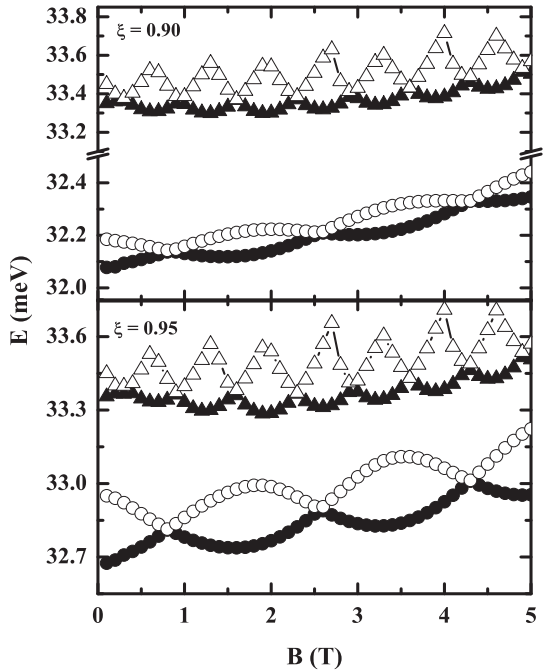}
{\caption{ Energy spectrum as a function of the magnetic field for
electrons in a double ring system, consisting of a circular
outer ring with $R_{1,e} = 400$ \AA\, and $R_{2,e} = 500$
\AA\,, and an elliptic inner ring considering $\xi = 0.90$
(top) and $\xi = 0.95$ (bottom).} }
\end{figure}

Figure 2(a) shows the confinement energies of the four low-lying
electron states in circular double rings as a function of the
magnetic field. The inner ring is limited by internal and
external radii $R_{1,i} = 230$ \AA\, and $R_{2,i} = 330$ \AA\,,
respectively, whereas the outer ring has radii $R_{1,e} = 400$
\AA\, and $R_{2,e} = 500$ \AA\,. The dependence of the energy spectrum on the different rings widths was already studied in previous papers, \cite{Li, Xiaojing} therefore, in this work, we restrict ourselves to the analysis of the specific case of concentric rings with the same width. As the magnetic field strength increases, the energies exhibit non periodic oscillations, which
differs from the energy spectrum of single rings reported in the
literature, where periodic AB oscillations have been found
\cite{9,11}. When both inner and outer rings exhibit
circular symmetry, as in the present case, the energy spectrum of
the double ring system can be obtained analytically. \cite{Latge1}
Indeed, the numerically obtained energy spectrum in Fig. 2 is
comparable to the analytical results presented in Fig. 4 of Ref.
\cite{Latge1} or, equivalently Fig. 1 of Ref.
\cite{Latge2}. As we mentioned, in the system that we have
considered, both the inner and outer rings have the same
width, so that the eigenfunctions have no preferential ring to be
confined in. Indeed, in the absence of tunnel coupling, each ring
spans an energy spectrum that is periodic in $B$ with different
periods according to Eq.(7), which are represented by the open and full symbols in the inset of Fig. 2, which illustrates the energy spectrum for two concentric rings delimited by a high potential barrier, i.e. with negligible coupling. The inter-ring coupling admix the two series of levels and destroys the periodic AB oscillations.
Correspondingly, the resulting eigenstates delocalize between the
two rings near an inter-ring resonance.  This effect is illustrated
in Fig. 2(b), which shows the calculated average radius $\langle
\Psi_0 | r |\Psi_0\rangle$ of the ground state. The average radius starts at the
barrier region for $B = 0$, showing that in the absence of the magnetic field, the ground
state is indeed distributed between the two rings. As $B$ increases, the average radius
jumps between the inner and outer rings, leading to the non-periodic energy oscillation at low
fields. However, for $B \gtrsim 2.6$, $\langle
\Psi_0 | r |\Psi_0\rangle$ stabilizes at the inner ring and, consequently, the ground state energy starts to oscillate periodically with $B$.

\begin{figure} [!h]
\centering
\includegraphics[width= 0.8\linewidth]{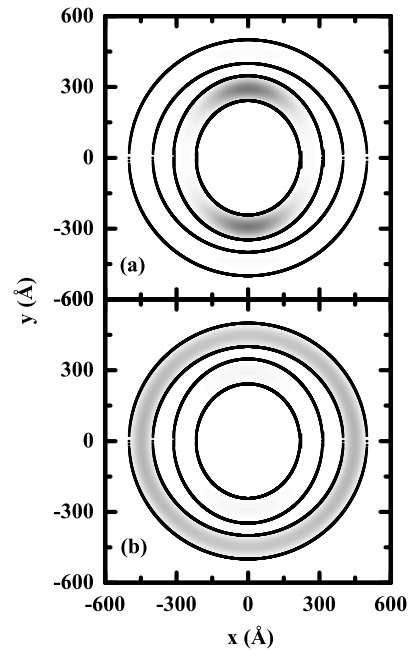}
{\caption{Contour plots of the (a) ground ($\Psi_0$) and (b)
second excited ($\Psi_2$) state eigenfunctions for electrons
confined in a system consisting of a circular outer ring with
$R_{1,e} = 400$ \AA\, and $R_{2,e} = 500$ \AA\,, and an elliptic
inner ring with $\xi = 0.90$, in the absence of magnetic
fields. The solid lines represent the limits of the rings.}}
\end{figure}

The previous results were obtained for two QRs with circular
basis. However, as mentioned above, the eigenstates of single QRs
are very sensitive to disorder, which are expected in actual
samples.  It is thus worth inquiring about the role of potential
anisotropy on coupled QRs.  To this end, we consider in the
following the effect of a small deviation from a circular shape
for one of the interacting rings. The electron energy spectrum for
a system consisting of an elliptic inner ring and a circular
outer ring is shown in Fig. 3. The inner ring presents
eccentricities $\xi = 0.90$ (top) and $\xi = 0.95$ (bottom) (its
average diameter is the same as the one in Fig. 1(a); see Ref.
\cite{Gil}). In this case, as only one of the rings is
elliptic, it is hard to write the Hamiltonian in circular
coordinates and find an analytical solution. More generally,
analytical solutions cannot be obtained for arbitrary perturbation
to the double quantum ring potential. Our method, however, allows
dealing with defects of arbitrary shape and strength.

\begin{figure} [!h]
\centering
\includegraphics[width= 1.0\linewidth]{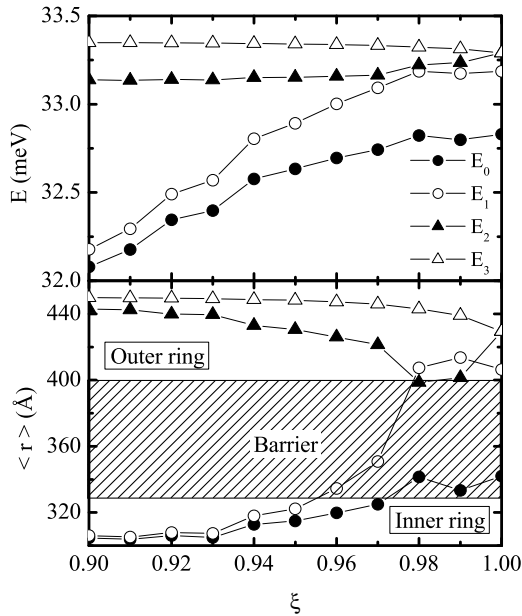}
{\caption{(a) Energies $E_j$ and (b) average radii
$\langle r \rangle = \langle \Psi_j | r |\Psi_j\rangle$ of the four low-lying electron states as a
function of the eccentricity of the inner ring.}}
\end{figure}

The results in Fig. 3 show that as the eccentricity of the inner ring
decreases, the ground ($E_0$) and first excited ($E_1$) states
become more energetically separated from the second ($E_2$) and
third excited ($E_3$) states. It can be clearly seen that the AB
oscillations were recovered by the existence of an elliptic ring.
Besides, the periodicity of the second and third excited states
oscillations is different from the one found for the ground and
first excited states. As the AB oscillations period is inversely
proportional to the ring area, the higher period found for $E_0$
and $E_1$ oscillations indicates that these states are confined at
the inner ring, while states corresponding to the energies
$E_2$ and $E_3$ are then confined at the outer ring. Indeed,
the period of oscillation for $E_0$ and $E_1$ ($E_2$ and $E_3$) is
practically insensitive to $\xi$, as obtained for a single QR
\cite{Gil}, and equal to $\Delta B \approx 1.75$ T (0.7 T),
corresponding to AB oscillations for an electron confined in a $R
= \sqrt{\phi_0\big/\pi\Delta B} \approx 274 $\AA\, (433 \AA\,)
quantum ring, according to Eq. (\ref{fig:ABspectrum}), which is
inside the inner (outer) ring region. The eigenfunctions $\Psi_0$
and $\Psi_2$ for the $\xi = 0.90$ case are shown in Figs. 4(a) and
4(b), respectively, where we confirm that the former is confined
in the inner (circular) ring, whereas the latter is confined in
the outer (elliptic) ring. Similarly, the eigenfunctions
$\Psi_1$ and $\Psi_3$ are localized within the inner and outer
rings, respectively (not shown).

Notice that even for deviations in the circular ring geometry as small as those shown in Figs. 3 and 4, the energy states in the inner and outer rings are already decoupled. It is then interesting to analyze how the coupled states in the circular case ($\xi = 1$) decouple as the eccentricity of the inner ring decreases from 1. To this purpose, in Fig. 5 we show the energy spectrum (a) and the average radii (b) of the four low-lying electron states as a function of the eccentricity of the inner ring. We observe that decreasing $\xi$ breaks the degeneracy of the $E_2$ and $E_3$ states in the absence of magnetic fields and leads to a red shift of $E_0$ and $E_1$, which enhances the gap between these two pairs of states. Moreover, the results in Fig. 5(b) demonstrate that in the circular double ring case, the ground state is confined in the barrier region, as mentioned above, but the higher energy states are confined in the outer ring. However, as $\xi$ decreases, the average radius of the first excited state decreases and eventually reaches the inner ring region for $\xi \leq 0.95$. For smaller values of $\xi$, the average radii of the ground and first excited (second and third excited) states converge to a value inside the inner (outer) ring region, as discussed earlier.

Finally, it is interesting to note that the two circular rings are strongly
tunnel coupled, as evidenced by the strongly deformed energy
spectrum in Fig. 2, as regards the spectrum we would obtain by a
simple addition of the spectra of two independent rings. On the other hand, Fig. 3
shows that the eccentricity tends to decouple the two QRs.  This
result is not straightforward, since the eccentricity brings into
play two opposite effects. Indeed, it introduces a region of
higher curvature, which leads to an angular localization of the
inner-related states.  Such localized states are expected to less
strongly interact with the more delocalized states of the outer
ring (see Fig. 4).  However, the inter-ring barrier potential is
thinner in the angular region where the ground inner state is
localized.  This would on the contrary enhance the inter-ring
interaction.  The results in Figs. 3-5 show that the first effect
dominates in the structure we consider here: even though the
eccentricity locally favors inter-QR coupling by a decrease of the
tunnel barrier, it also render the coupling non-resonant by
introducing a red-shift of the localized inner states that
increases with decreasing $\xi$, as shown in Fig. 5.

The results presented in this paper regard only to one kind of perturbation of the geometry of the inner ring, namely, to deviations in the inner ring eccentricity. In this case, we demonstrated that the ground and first excited states are trapped in the higher curvature regions of the inner ring, decoupling inner and outer ring states. Other kinds of geometry perturbation in the inner ring are possible, which would produce regions with higher curvature \cite{Gridin, Alfonso} or larger width \cite{Wanderley} in this ring. In both cases, these regions would be more energetically favorable for the electrons confinement and thus, we expect that they would also be able to decouple the inner and outer ring states, otherwise coupled in the case of perfectly circular rings with the same width. However, the confirmation of this effect needs further investigation, and is left for future works.

\section{Conclusions}

A theoretical investigation of the electron confinement in double
concentric quantum rings under applied magnetic fields was
performed, for circular and elliptic rings. When circular rings
are considered, the energy spectrum exhibits non-periodic
oscillations as the magnetic field increases. However, periodic AB
oscillations are found when one considers an elliptic inner
ring, even when one deals with small eccentricities. Moreover,
when one analyzes the periodicity of these AB oscillations, it can
be seen that the eigenstates in such a system are spatially
separated: ground and first excited states are confined at the
elliptic inner ring, whereas the second and third excited
states are confined at the circular outer ring.

\bigskip
\noindent
\textbf{Acknowledgments}
\bigskip

We thank F. M. Peeters for a critical reading of the manuscript. This work has received financial support from the Brazilian
National Research Council (CNPq), under contract NanoBioEstruturas
555183/2005-0, CAPES and PRONEX/CNPq/FUNCAP.


\begin{thebibliography}{99}
\bibitem{Ledentsov} N. N. Ledentsov, V. M. Ustinov, V. A. Shchukin, P. S. Kop'ev, Zh. I. Alferov, and D. Bimberg, Semiconductors \textbf{32}, 343 (1998).
\bibitem{Skolnick} M.S. Skolnick and D.J. Mowbray, Annu. Rev. Mater. Res. \textbf{34}, 181 (2004).
\bibitem{Kleemans} N. A. J. M. Kleemans, I. M. A. Bominaar-Silkens, V. M. Fomin, V. N. Gladilin, D. Granados, A. G. Taboada, J. M. Garc\'ia, P. Offermans, U. Zeitler, P. C. M. Christianen, J. C. Maan, J. T. Devreese, and P. M. Koenraad, Phys. Rev. Lett. \textbf{99}, 146808 (2007).
\bibitem{Warburton} R. J. Warburton, C. Schulhauser, D. Haft, C. Sch\"aflein, K. Karrai, J. M. Garc\'ia, W. Schoenfeld, and P. M. Petroff, Phys. Rev. B \textbf{65}, 113303 (2002).
\bibitem{7} D. Granados and J. M. Garc\'ia, Appl. Phys. Lett. {\bf 82},
2401 (2003).
\bibitem{Aharonov} Y. Aharonov and D. Bohm, Phys. Rev. \textbf{115}, 485 (1959).
\bibitem{Wendler} L. Wendler, V. M. Fomin, and A. V. Chaplik, Solid State Commun. \textbf{96}, 809 (1995).
\bibitem{1} J. Planelles and J. I. Climente, Eur. Phys. J. B {\bf 48}, 65 (2005).
\bibitem{6} A. V. Maslov and D. S. Citrin, Phys. Rev. B {\bf 67}, 121304(R) (2003).
\bibitem{5} T. Kuroda, T. Mano, T. Ochiai, S. Sanguinetti, T.
Noda, K. Kuroda, K. Sakoda, G. Kido and N. Koguchi, Physica E {\bf
32}, 46 (2006).
\bibitem{2} T. Mano, T. Kuroda, S. Sanguinetti, T. Ochiai, T. Tateno, J.
Kim, T. Noda, M. Kawabe, K. Sakoda, G. Kido and N. Koguchi, Nano
Lett. {\bf 5}, 425 (2005).
\bibitem{Sanguinetti} S. Sanguinetti, M. Abbarchi, A. Vinattieri, M. Zamfirescu, M. Gurioli, T. Mano, T. Kuroda, and N. Koguchi, Phys. Rev. B \textbf{77}, 125404 (2008).
\bibitem{Latge1} F. J. Culchac, N. Porras-Montenegro, J. C. Granada, and
A. Latg\'e, Microelectron. J. \textbf{39}, 402 (2008).
\bibitem{Latge2} F. J. Culchac, N. Porras-Montenegro, and A. Latg\'e, J. Phys.: Condens. Matter \textbf{20}, 285215
(2008).
\bibitem{11} B. Szafran and F. M. Peeters, Phys. Rev. B {\bf 72}, 155316
(2005).
\bibitem{Latge3} F. J. Culchac, N. Porras-Montenegro, and A.
Latg\'e, J. Appl. Phys. \textbf{105}, 094324 (2009).
\bibitem{Planelles2} J. Planelles and J.I. Climente, Eur. Phys. J. B \textbf{48}, 65
(2005).
\bibitem{Vanska} T. V\"ansk\"a and D. Sundholm, Phys. Rev. B \textbf{82}, 085306 (2010).
\bibitem{Gridin} D. Gridin, A. T. I. Adamou, and R. V. Craster, Phys. Rev. B \textbf{69}, 155317
(2004).
\bibitem{Alfonso} A. Bruno-Alfonso and A. Latg\'e, Phys. Rev. B \textbf{77}, 205303 (2008).
\bibitem{Gil} G. A. Farias, M. H. Degani, J. A. K. Freire, J. Costa e Silva, and R.
Ferreira, Phys. Rev. B \textbf{77}, 085316 (2008).
\bibitem{Planelles} J Planelles,, F Rajadell, and J I Climente, Nanotechnology \textbf{18}, 375402
(2007).
\bibitem{3} F. Palmero, J. Dorignac, J. C. Eilbeck and R. A. R\"omer, Phys. Rev. B {\bf 72}, 075343 (2005).
\bibitem{8} A. Bruno-Alfonso and A. Latg\'e, Phys. Rev. B {\bf 71}, 125312
(2005).
\bibitem{meu} A. Chaves, G. A. Farias, F. M. Peeters, and B.
Szafran, Phys. Rev. B 80, 125331 (2009).
\bibitem{4} M. H. Degani, Phys. Rev. B {\bf 66}, 233306 (2002).
\bibitem{10} J. Costa e Silva, A. Chaves, J. A. K. Freire, V. N. Freire and G. A. Farias,
 Phys. Rev. B {\bf 74}, 085317 (2006).
\bibitem{Li} S.-S. Li and J.-B. Xia, Nanoscale Res. Lett. \textbf{1}, 167 (2006).
\bibitem{Xiaojing} X. Li, Physica E \textbf{41}, 1814 (2009).
\bibitem{9} L. G. G. V. Dias da Silva, S. E. Ulloa and A. O.
Govorov, Phys. Rev. B {\bf 70}, 155318 (2004).
\bibitem{Wanderley} L. A. Lavenere-Wanderley, A. Bruno-Alfonso, and A. Latg\'e, J. Phys.: Condens. Matter \textbf{14}, 259 (2002).

\end{thebibliography}
\end{document}